# When to Crossover from Earth to Space for Lower Latency Data Communications?

Aizaz U. Chaudhry, *Senior Member, IEEE,* and Halim Yanikomeroglu, *Fellow, IEEE*

*Abstract*—For data communications over long distances, optical wireless satellite networks (OWSNs) can offer lower latency than optical fiber terrestrial networks (OFTNs). However, when is it beneficial to switch or crossover from an OFTN to an OWSN for lower latency data communications? In this work, we introduce a crossover function that enables to find the crossover distance, i.e., a distance between two points on the surface of the Earth beyond which switching or crossing over from an OFTN to an OWSN for data communications between these points is useful in terms of latency. Numerical results reveal that a higher refractive index of optical fiber (or $i$) in an OFTN and a lower altitude of satellites (or $h$) in an OWSN result in a shorter crossover distance. To account for the variation in the end-to-end propagation distance that occurs over the OWSN, we examine the crossover function in four different scenarios. Numerical results indicate that the crossover distance varies with the end-to-end propagation distance over an OWSN and is different for different scenarios. We calculate the average crossover distance over all scenarios for different $h$ and $i$ and use it to evaluate the simulation results. Furthermore, for a comparative analysis of OFTNs and OWSNs in terms of latency, we study three different OFTNs having different refractive indices and three different OWSNs having different satellite altitudes in three different scenarios for long-distance inter-continental data communications, including connections between New York and Dublin, Sao Paulo and London, and Toronto and Sydney. All three OWSNs offer better latency than OFTN2 (with $i_2 = 1.3$) and OFTN3 (with $i_3 = 1.4675$) in all scenarios. For example, for Toronto–Sydney connection, OWSN1 (with $h_1 = 300$ km), OWSN2 (with $h_2 = 550$ km) and OWSN3 (with $h_3 = 1,100$ km) perform better than OFTN2 by 18.11%, 16.08%, and 10.30%, respectively, while they provide an improvement in latency of 27.46%, 25.67%, and 20.54%, respectively, compared to OFTN3. The OWSN1 performs better than OFTN1 (with $i_1 = 1.1$) for Sao Paulo–London and Toronto–Sydney connections by 2.23% and 3.22%, respectively, while OWSN2 outperforms OFTN1 for Toronto–Sydney connection by 0.82%. For New York–Dublin connection, all OWSNs while for Sao Paulo–London connection, OWSN2 and OWSN3 exhibit higher latency than OFTN1 as the corresponding average crossover distances are greater than the shortest terrestrial distances between cities in these scenarios. Multiple satellites (or laser inter-satellite links) on its shortest paths drive up the propagation distance to the extent that OWSN3 ends up with a higher latency than OFTN1 for the Toronto–Sydney inter-continental connection scenario although the related average crossover distance is less than the shortest terrestrial distance between Toronto and Sydney. The challenges related to OWSNs and OFTNs that may arise from this work in future are also highlighted.

*Index Terms*—Crossover distance, crossover function, LEO, network latency, optical fiber terrestrial networks, optical wireless satellite networks, satellite constellations.

A.U. Chaudhry and H. Yanikomeroglu are with the Department of Systems and Computer Engineering, Carleton University, Ottawa, ON, Canada – K1S 5B6, e-mail: {auhchaud, halim}@sce.carleton.ca.

## I. INTRODUCTION

OPTICAL wireless satellite networks (OWSNs), also known as free-space optical satellite networks, will be created in space by employing laser inter-satellite links (LISLs) between satellites in upcoming low Earth orbit (LEO) or very low Earth orbit (VLEO) satellite constellations, like SpaceX's Starlink [1] and Telesat's Lightspeed [2]. The LISLs [3], also known as optical inter-satellite links, will be essential in ensuring low-latency paths (or routes) within the OWSN [4], [5]. Without LISLs, a long-distance inter-continental data communications connection between two cities, such as New York and Dublin, will have to bounce up and down between ground stations and satellites, and this will negatively affect latency of the satellite network.

The provision of long-distance low-latency data communications could be the primary use case of an OWSN that is created by using LISLs in an upcoming LEO/VLEO satellite constellation like Phase I of Starlink [6]. An OWSN can offer such type of communications as a premium service to the financial hubs around the globe, and such a use-case can be beneficial in recovering the cost of deploying and sustaining such satellite networks. In high-frequency trading (HFT) of stocks at the stock exchange, a one millisecond advantage can translate into $100 million a year in revenues for a major brokerage firm [7]. An advantage of a few milliseconds in HFT could mean billions of dollars of revenues for these firms. Technological solutions in the form of communications networks that can provide lower latency data communications are being highly coveted by such firms, and a low-latency OWSN could be the ideal solution.

Unlike optical fiber terrestrial networks (OFTNs) where data is sent using a laser beam over a guided medium, i.e., optical fiber, data communications in OWSNs takes places over LISLs by using laser beams between satellites over an unguided medium, i.e., vacuum of space. The refractive index of a medium indicates the speed of light through that medium, and a higher refractive index means a slower transmission of light [8]. Optical fibers typically have a refractive index of approximately 1.5. This means that the speed of light in an optical fiber is approximately $c/1.5$, where $c$ is the speed of light in vacuum [9]. This translates into speed of light in vacuum being approximately 50% higher than the speed of light in optical fiber. This has significant implications, and the higher speed of light in OWSNs operating in the vacuum of space gives them a critical advantage over OFTNs in terms of latency for long-distance data communications.

The delay from source to destination in the network is composed of transmission delay, processing delay, queueing delay, and propagation delay [10]. For optical communications in OFTNs or OWSNs over optical fiber or vacuum of space, propagation delay is the delay arising from the transmission of the optical signal along the medium. It is directly proportional to the distance between the source and the destination and becomes very significant in long-distance data communications [11]. In this work, we study the latency of OWSNs and OFTNs, and we define latency (or end-to-end network latency) as the propagation delay from the source to the destination.

Firstly, we propose a crossover function for data communications between two points on the surface of the Earth over an OFTN and an OWSN. The crossover function is then used to calculate the crossover distance that indicates when switching or crossing over from an OFTN to an OWSN can be beneficial for data communications in terms of latency. The crossover function and thereby the crossover distance depend upon the refractive index of the optical fiber in an OFTN and the altitude of satellites in an OWSN. From numerical results, we observe that a higher optical fiber refractive index and a lower altitude of satellites result in a shorter crossover distance.

We examine crossover function in four different scenarios to account for the different end-to-end propagation distances that occur over the OWSN due to the orbital movement of satellites with time. The end-to-end propagation distance over an OWSN is smallest in Scenario 3, increases in Scenario 1, increases further in Scenario 4, and is largest in Scenario 2. The numerical results show that the crossover distance varies with the end-to-end propagation distance over an OWSN, it is minimum for Scenario 3, it is higher for Scenario 1 as compared to that for Scenario 3, it is more for Scenario 4 than that for Scenario 1, and it is maximum for Scenario 2. We calculate the average crossover distance over all scenarios for different altitudes of satellites and different refractive indices. We later use the average crossover distance to evaluate the simulation results.

Next, we investigate the impact on latency of different optical fiber refractive indices in OFTNs and different altitudes of satellites in OWSNs and conduct a comparative analysis of these networks. To this end, we consider three different OFTNs with different refractive indices and three different OWSNs having different satellite altitudes, and compare them in terms of latency under three different scenarios for long-distance inter-continental data communications, including connections between New York and Dublin, Sao Paulo and London, and Toronto and Sydney. For this comparison, we use Starlink's Phase I constellation and LISLs between satellites to simulate an OWSN. Using three different altitudes of satellites for this constellation including 300 km, 550 km, and 1,100 km, we simulate the three different OWSNs. We consider 1.1, 1.3, and 1.4675 as the refractive indices for the three different OFTNs. Note that 1.4675 [12] is assumed as the refractive index for one of these OFTNs since it is the refractive index of existing long-haul submarine optical fiber cables that provide inter-continental connectivity. Optical fiber refractive indices may be reduced in future due to advancements in the optical fiber technology and thereby we consider OFTNs with lower refractive indices of 1.3 and 1.1 as well.

For each scenario, we find minimum-latency paths between cities over these six different networks. From the results, we observe that all three OWSNs outperform the OFTN with a realistic refractive index of 1.4675 as well as the OFTN with a refractive index of 1.3 in terms of latency in all scenarios. The OWSN operating at 300 km altitude performs closely to the OFTN with a refractive index of 1.1 for the New York–Dublin connection and outperforms it for the Sao Paulo–London and Toronto–Sydney connections while the OWSN at 550 km altitude offers lower latency than this OFTN for the Toronto–Sydney connection. Within all scenarios, it is observed that the lower the refractive index of an OFTN or the lower the altitude of satellites in an OWSN, the lower the latency of a network. For data communications in different scenarios over a network, it is seen that the greater the inter-continental distance between cities, the higher the latency of a network. Preliminary work in this regard has appeared in [13].

Unlike LISLs that are established between satellites in the vacuum of space, laser uplink (i.e., laser link between a ground station and a satellite) and laser downlink (i.e., laser link between a satellite and a ground station) communication suffers from atmospheric attenuation, such as Mie scattering and geometrical scattering [14], as the laser beam propagates through Earth's atmosphere. This attenuation affects the performance of uplink and downlink communication and can even disrupt this communication on days when the line-of-sight between a ground station and the overhead satellite is blocked by cloudy weather. A possible solution to deal with link outage caused by adverse weather conditions is site diversity, which involves selecting the best ground station that provides the best channel conditions [15]. Another possible solution, referred to as hybrid radio frequency and free-space optical communication, consists of using radio frequency link in combination with free-space optical (or laser) link and selecting one of them for uplink and downlink communication depending on the weather conditions [16]. Note that the focus of the work in our paper is on latency in OFTNs and OWSNs and investigating link outage in laser uplink and laser downlink communication and its possible solutions are out of the scope of this work.

The rest of the paper is organized as follows. The related work along with some use cases and the motivation of this work are discussed in Section II. Section III presents crossover functions for the different scenarios and the related numerical results. Our methodology for calculating latency of an inter-continental connection between cities over an OFTN and an OWSN is given in Section IV. Section V presents results for the comparison of different OFTNs and different OWSNs in terms of latency. Conclusions are summarized in Section VI. Section VII highlights future challenges.

## II. Motivation

In addition to providing an ideal solution for low-latency long-distance inter-continental data communications for HFT between stock exchanges around the globe, OWSNs can also be beneficial in other scenarios. In one such use case, an OWSN can help in extending broadband Internet to rural and remote areas when integrated as a backbone with the existing 4G/5G networks. The OWSN can provide a backhaul network to connect the 4G/5G access networks in rural and remote areas to their core network. By providing the ability to connect any two points on Earth over the OWSN arising from employing LISLs between satellites in their upcoming satellite constellation Lightspeed, Telesat plans to enable the provision of broadband Internet to unserved and underserved communities and individuals in rural and remote areas. Telesat has already successfully demonstrated this use case for Internet backhauling with TIM Brasil – an Internet Service Provider – in their 4G network where a Lightspeed's Phase I satellite was used to connect remote communities in Brazil to the Internet by linking TIM Brasil's access network to its core [17].

In another use case for OWSNs, the European Space Agency's High thRoughput Optical Network (HydRON) project is targeting a "Fiber in the Sky" network in space [18]. The goal of this project is to enable an all-optical transport network in space. It will utilize all-optical payloads interconnected via Tbps optical inter-satellite links to realize a true "Fiber in the Sky" network. The HydRON system is expected to have the following main functionalities: bidirectional very high-capacity laser inter-satellite links and reliable very high-capacity optical feeder links; interface compatibility with RF/optical customer payloads for traffic distribution/collection to/from these payloads; on-board fast transparent optical switching and on-board fast regenerative electrical switching; and network optimization using artificial intelligence [19].

Developing a better understanding of latency in satellite networks arising from upcoming LEO/VLEO satellite constellations has been the focus of the research community [4], [5], [13], [20]–[27]. A study has investigated the use of ground-based relays as a substitute of inter-satellite links to provide low-latency communications over satellite networks [4]. It is concluded that lower latency is achieved when inter-satellite links are used between satellites in a satellite constellation at 550 km altitude than using ground stations as relays. It is reported in a study that inter-satellite links substantially reduce latency variations in the satellite network [5]. A simulator for studying different parameters, including latency, of satellite networks arising from upcoming LEO/VLEO satellite constellations has been developed [20]. It is stated in [21] that a satellite network using inter-satellite links within a satellite constellation can provide lower latency than an OFTN for data communications over long distances greater than 3,000 km.

It is mentioned that a dense small satellite network (i.e., a satellite network arising from a satellite constellation consisting of hundreds of satellites, such as Starlink and Lightspeed) has the potential to provide lower latency than any terrestrial network of comparable length due to the higher speed of light in free space than in optical fiber [22]. It is further stated that although the target latency of 1 ms for 5G cellular systems cannot be directly attained with the help of a dense small satellite network, it may indirectly support 5G networks in decreasing latency by offering alternate backhaul. Considering the very narrow beam divergence of optical communications between satellites, the architecture of two different satellite networks consisting of 120 and 1,600 satellites, respectively, is evaluated with respect to satellite antenna steering capability and satellite visibility [23]. Guidelines are given for designing a visibility matrix for the time-varying satellite topology. The proposed approach shows better performance than the classical approach of pre-assigned links in terms of end-to-end link distance.

It is noted that the low altitude of LEO satellites compared to medium Earth orbit and geostationary Earth orbit satellites enables users to experience a round-trip delay that is similar to the one on terrestrial links [24]. It is also mentioned that due to the ultra-dense topology of a LEO satellite network, a terrestrial satellite terminal has multiple LEO satellites to choose from and tends to select the shortest terrestrial satellite terminal-to-satellite link to reduce the propagation delay for the access part of the satellite network. The network delay has been analyzed in a multihop satellite network, where the satellites are modelled as a queue network connected in series [25]. The network delay is modelled as a function of the system utilization load, and it is observed that the network delay increases with the system utilization load. Insights are provided for designing multihop satellite networks for latency-sensitive applications.

A study considers a hypothetical constellation of 1,600 LEO satellites in 40 orbital planes with 40 satellites in each orbital plane at 550 km altitude and 53° inclination, and shows a median round trip time improvement of 70% with the satellite network based on this constellation when comparing with Internet latency [26]. However, this cannot be considered a fair comparison as it overly favors the satellite network. Delays due to sub-optimal routing, congestion, queueing, and forward error correction are not considered in the satellite network while such delays are counted in measuring Internet latency. In our work, a comparison of an OWSN and an OFTN may be favorable to the OFTN. We consider the shortest distance between two cities over an OFTN along the Earth's surface, which is not the case in reality. An OFTN cannot provide the shortest distance path along surface of the Earth between cities in different continents. Note that long-haul submarine optical fiber cables are laid along paths that avoid earthquake prone areas and difficult seabed terrains with high slopes instead of following the shortest path to connect two points on Earth's surface [28].

In an earlier work on the use of free-space optics for next-generation satellite networks, we investigate a use case that checks the suitability of OWSNs for providing low-latency communications over long distances as their primary service [27]. It is shown that an OWSN operating at 550 km altitude can outperform an OFTN in terms of latency when data communications takes place over distances that are greater than 3,000 km. In preliminary work [13], we built on our work in [27] and studied the latency of OWSNs versus OFTNs in realistic scenarios. Unlike any previous work in the literature, we introduce a novel crossover function in this work and subsequently use it to calculate the crossover distance, i.e., a parameter which indicates when to crossover from an OFTN on Earth to an OWSN in space for lower latency data communications. Furthermore, we study the impact on latency of different optical fiber refractive indices in OFTNs and different altitudes of satellites in OWSNs in different scenarios for long-distance inter-continental data communications and conduct a comparative analysis of these networks in terms of latency. To the best of our knowledge, such a study does not exist in the literature.

## III. CROSSOVER FUNCTION

The crossover function can be defined as a function that can be used towards deciding whether to send the data traffic over the OFTN or the OWSN, i.e., when switching or crossing over from one network to the other is beneficial for data communications in terms of latency. In the following, we present this crossover function in four different scenarios.

### A. Scenario 1

Let us consider the first scenario (i.e., Scenario 1) shown in Fig. 1, where the Earth is shown in blue color, two points $A$ and $B$ on the surface of the Earth are shown in green color and two LEO satellites $X$ and $Y$ at altitude $h$ are shown in yellow color. Let us assume that $A$ and $B$ can communicate over optical fiber along the Earth's surface in the OFTN and these points also have the option to communicate using satellites $X$ and $Y$ in the OWSN.

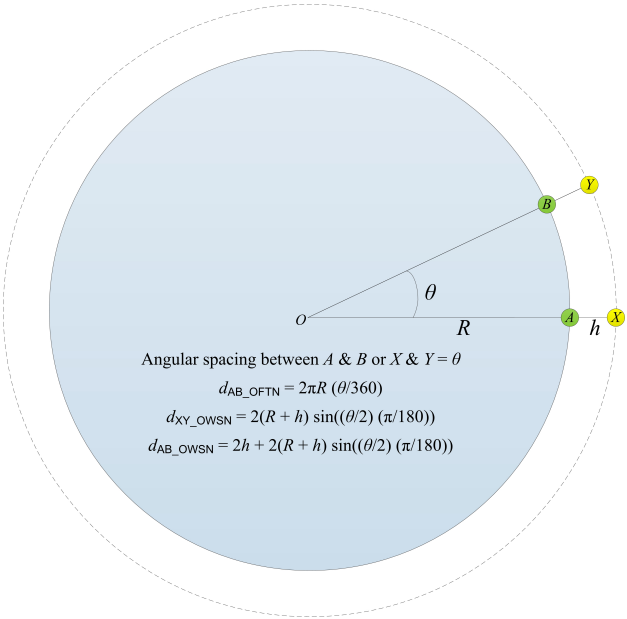

Fig. 1. Scenario 1 illustrating end-to-end propagation distance between $A$ and $B$ over the OFTN and the OWSN. Earth is shown in blue color, points $A$ and $B$ (which can be optical fiber relay stations in the OFTN or ground stations in the OWSN) on the surface of the Earth are shown in green, and LEO satellites $X$ and $Y$ in space are shown in yellow. The dashed circle indicates the orbit of the satellites and the point $O$ represents the center of the Earth.

Let $\theta$ be the angular spacing in degrees between $A$ and $B$ and between $X$ and $Y$. The distance between $O$ and $A$ is equal to the radius of the Earth and is denoted as $R$. The distance between $O$ and $X$ is equal to $R + h$ and is denoted as $r$.

For the OFTN, the end-to-end propagation distance between $A$ and $B$ on this network is equal to the length of the optical fiber along the surface of the Earth between $A$ and $B$. This is equal to the length of the arc $AB$ and is given by

$$d_{\text{AB\_OFTN}} = 2\pi R \left(\frac{\theta}{360}\right). \quad (1)$$

Note that a LISL between two satellites in the OWSN propagates in a straight line between these satellites and not along an arc.

Unlike the OFTN where the propagation distance is calculated along an arc, the propagation distance between $X$ and $Y$ on the OWSN is length of the straight line between these satellites. This is equal to the length of the chord $XY$ and is given by

$$d_{\text{XY\_OWSN}} = 2r \sin\left(\left(\frac{\theta}{2}\right)\left(\frac{\pi}{180}\right)\right). \quad (2)$$

Substituting $r = R + h$ in (2), we get

$$d_{\text{XY\_OWSN}} = 2(R+h) \sin\left(\left(\frac{\theta}{2}\right)\left(\frac{\pi}{180}\right)\right). \quad (3)$$

When the data communications between $A$ and $B$ takes place over the OWSN, this incurs an extra propagation distance $h$ for uplink (i.e., the distance between $A$ and satellite $X$) as well as an extra propagation distance $h$ for downlink (i.e., the distance between satellite $Y$ and $B$). Now the end-to-end propagation distance between $A$ and $B$ for data communications over the OWSN in Scenario 1 can be calculated as

$$d_{\text{AB\_OWSN}} = 2h + 2(R+h) \sin\left(\left(\frac{\theta}{2}\right)\left(\frac{\pi}{180}\right)\right). \quad (4)$$

Let $i$ be the refractive index of optical fiber in the OFTN and $c$ be the speed of light in vacuum. Then, the end-to-end latency (or propagation delay) for the end-to-end propagation distance $d_{\text{AB\_OFTN}}$ over the OFTN is given by

$$t_{\text{AB\_OFTN}} = d_{\text{AB\_OFTN}} \left(\frac{i}{c}\right). \quad (5)$$

Substituting $d_{\text{AB\_OFTN}}$ in (5), we get

$$t_{\text{AB\_OFTN}} = 2\pi R \left(\frac{\theta}{360}\right)\left(\frac{i}{c}\right). \quad (6)$$

Similarly, the end-to-end latency for the end-to-end propagation distance $d_{\text{AB\_OWSN}}$ over the OWSN is given by

$$t_{\text{AB\_OWSN}} = \frac{d_{\text{AB\_OWSN}}}{c}. \quad (7)$$

Substituting $d_{\text{AB\_OWSN}}$ in (7), we get

$$t_{\text{AB\_OWSN}} = \frac{2h + 2(R+h) \sin\left(\left(\frac{\theta}{2}\right)\left(\frac{\pi}{180}\right)\right)}{c}. \quad (8)$$

Crossover function is the ratio of the end-to-end latencies over the two networks and is given by

$$f_{crossover}(\theta) = \frac{t_{\text{AB\_OWSN}}}{t_{\text{AB\_OFTN}}}. \quad (9)$$

Substituting $t_{\text{AB\_OWSN}}$ and $t_{\text{AB\_OFTN}}$ in (9) and after cancelling $c$ in the numerator and the denominator, we get crossover function for Scenario 1 as

$$f_{1\_crossover}(\theta) = \frac{2h + 2(R+h) \sin\left(\left(\frac{\theta}{2}\right)\left(\frac{\pi}{180}\right)\right)}{2\pi R \left(\frac{\theta}{360}\right)(i)}. \quad (10)$$

Let $r_{gs}$ denote the range of ground stations at points $A$ and $B$ and $\varepsilon$ represent the elevation angle between $A$ and satellite $X$ or $B$ and satellite $Y$. For a given elevation angle $\varepsilon$, the range of a ground station (also known as slant range between a ground station and a satellite) can be calculated using [29]

$$r_{gs} = R\left[\sqrt{\left(\frac{R+h}{R}\right)^2 - \cos^2(\varepsilon)} - \sin(\varepsilon)\right], \quad (11)$$

where $R$ is the radius of the Earth and $h$ is the altitude of the satellites in the constellation. The crossover function in (10) can then be written as

$$f_{1\_crossover}(\theta) = \frac{2r_{gs} + 2(R+h)\sin\left(\left(\frac{\theta}{2}\right)\left(\frac{\pi}{180}\right)\right)}{2\pi R\left(\frac{\theta}{360}\right)(i)}. \quad (12)$$

Substituting $r_{gs}$ in (12), we get (13). In this scenario, the satellites are assumed to be exactly over the ground stations in space, which means that $\varepsilon = 90°$, $r_{gs} = h$, and the crossover functions in (10) and (13) yield the same result.

The crossover function decreases monotonically with increase in $\theta$. Using (10), we calculate the value of $\theta$ where the crossover function is equal to 1. We call this value of $\theta$ as crossover $\theta$ or $\theta_{crossover}$. Putting the value of crossover $\theta$ in (1) gives us the value of the crossover distance or $d_{crossover}$. If the distance between $A$ and $B$ is greater than the crossover distance, then the latency of data communications between $A$ and $B$ over the OWSN will be less than that over the OFTN. This means that crossing over or switching to OWSN from OFTN is beneficial in terms of latency for data communications between two points on Earth when the distance between them is greater than the crossover distance.

Note that the value of crossover $\theta$ from (10) and hence the crossover distance from (1) depend upon the following two parameters:

- the altitude of the satellites in the OWSN or $h$, and
- the refractive index of the optical fiber in the OFTN or $i$.

Let us consider an example where we assume $h = 550$ km and $i = 1.5$. The radius of the Earth or $R$ is considered as 6,378 km. Using these values in (10), we get crossover $\theta$ or $\theta_{crossover}$ equal to 23.4533° and a plot of crossover function versus $\theta$ for this case is shown in Fig. 2. Substituting this value of $\theta_{crossover}$ in (1), we obtain the value of crossover distance equal to 2611 km for this scenario when $h = 550$ km and $i = 1.5$. This means that the OWSN operating at 550 km altitude with ingress and egress satellites at an elevation angle of 90° will have an advantage over the OFTN (consisting of optical fiber having a refractive index of 1.5) in terms of latency (or propagation delay) when data communications takes place between two points on Earth that are separated by a distance that is more than 2,611 km.

We calculate results for $\theta_{crossover}$ and $d_{crossover}$ for different values of $h$ and $i$ and these are given in Table 1. Note that $h$ and $r_{gs}$ are same in this scenario as shown in this table. We can see from these results that as the value of optical fiber refractive index $i$ decreases for some value of $h$, the values of the crossover parameters $\theta_{crossover}$ and $d_{crossover}$ increase. For example, at $h = 550$ km in Table 1, $d_{crossover}$ is 2,611 km, 2,820 km, 3,371 km, 4,632 km, 6,730 km, and 9,636 km at $i = 1.5, 1.4675, 1.4, 1.3, 1.2,$ and 1.1, respectively. The same trend is evident from Fig. 3 where crossover distance is plotted against $i$ at different values of $h$. This means that the smaller the $i$, the larger the crossover distance and the higher the distance between two points on Earth when data communications between them over an OWSN is more beneficial compared to an OFTN in terms of latency.

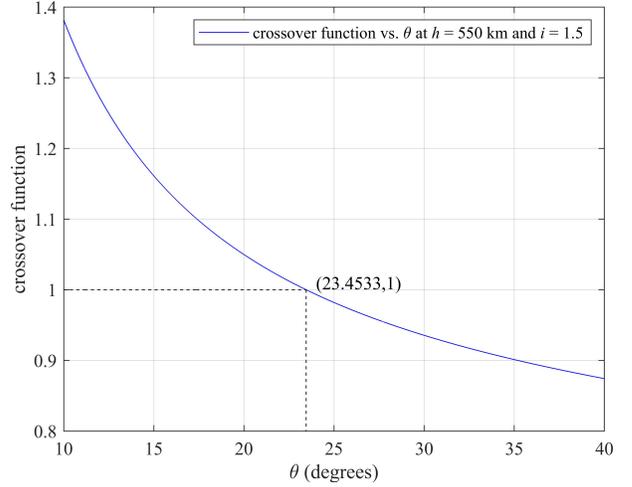

Fig. 2. Plot of crossover function for Scenario 1 vs. $\theta$ at $h = 550$ km and $i = 1.5$. The value of $\theta$ when the crossover function is equal to 1 is 23.4533, and we call it the crossover $\theta$. Putting this value of crossover $\theta$ in (1) gives the crossover distance of 2,611 km for Scenario 1 when $h = 550$ km and $i = 1.5$.

On the other hand, as the altitude of satellites or $h$ increases for some $i$, we can see from Table 1 that the crossover parameters increase indicating that the higher the $h$, the larger the crossover distance. For example, at $i = 1.4675$ in Table 1, $d_{crossover}$ is 1,420 km, 2,525 km, 2,820 km, 3,747 km, 5,055 km, and 6,402 km at $h = 300$ km, 500 km, 550 km, 700 km, 900 km, and 1,100 km, respectively. A plot of crossover distance against $h$ at different values of $i$ in Fig. 4 clearly illustrates this trend. In summary, we can say that a higher value of $i$ and a lower value of $h$ will result in a smaller value of the crossover distance which will be favorable to an OWSN in making it more efficient than an OFTN in terms of latency. On the other hand, a lower $i$ and a higher $h$ will favor the OFTN.

### B. Scenario 2

This scenario is illustrated in Fig. 5, where points $X'$ and $Y'$ indicate the positions of the satellites in Scenario 1, and yellow circles represent the positions of the satellites $X$ and $Y$ in this scenario. When assuming an anti-clockwise orbital movement of satellites, note that the LEO satellites $X$ and $Y$ are located before $X'$ and after $Y'$, respectively, in this scenario.

The end-to-end propagation distance over the OWSN in this scenario is equal to

$$d_{\text{AB\_OWSN}} = d_{AX} + d_{XX'} + d_{X'Y'} + d_{Y'Y} + d_{YB}, \quad (14)$$

where $d_{AX} = d_{YB} = r_{gs}$. The length of the chord $X'Y'$ in this scenario is the same as the length of the chord $XY$ in Scenario 1, i.e.,

$$d_{X'Y'} = 2(R+h)\sin\left(\left(\frac{\theta}{2}\right)\left(\frac{\pi}{180}\right)\right). \quad (15)$$

TABLE 1
SCENARIO 1 – CROSSOVER $\theta$ AND CROSSOVER DISTANCE FOR DIFFERENT VALUES OF $h$ AND $i$; $\varepsilon = 90°$.

| $h$ (km) | $i$ | $r_{gs}$ (km) | $\theta_{crossover}$ (degrees) | $d_{crossover}$ (km) |
|---|---|---|---|---|
| 300 | 1.5 | 300 | 11.8502 | 1,319 |
|  | 1.4675 |  | 12.7523 | 1,420 |
|  | 1.4 |  | 15.1405 | 1,685 |
|  | 1.3 |  | 20.8331 | 2,319 |
|  | 1.2 |  | 32.3167 | 3,597 |
|  | 1.1 |  | 56.6665 | 6,308 |
| 500 | 1.5 | 500 | 21.0061 | 2,338 |
|  | 1.4675 |  | 22.6813 | 2,525 |
|  | 1.4 |  | 27.0892 | 3,016 |
|  | 1.3 |  | 37.3393 | 4,157 |
|  | 1.2 |  | 55.1800 | 6,143 |
|  | 1.1 |  | 81.4951 | 9,072 |
| 550 | 1.5 | 550 | 23.4533 | 2,611 |
|  | 1.4675 |  | 25.3295 | 2,820 |
|  | 1.4 |  | 30.2780 | 3,371 |
|  | 1.3 |  | 41.6112 | 4,632 |
|  | 1.2 |  | 60.4530 | 6,730 |
|  | 1.1 |  | 86.5597 | 9,636 |
| 700 | 1.5 | 700 | 31.1414 | 3,467 |
|  | 1.4675 |  | 33.6584 | 3,747 |
|  | 1.4 |  | 40.1980 | 4,475 |
|  | 1.3 |  | 54.3524 | 6,050 |
|  | 1.2 |  | 74.9743 | 8,346 |
|  | 1.1 |  | 99.9820 | 11,130 |
| 900 | 1.5 | 900 | 42.0708 | 4,683 |
|  | 1.4675 |  | 45.4115 | 5,055 |
|  | 1.4 |  | 53.8290 | 5,992 |
|  | 1.3 |  | 70.4686 | 7,844 |
|  | 1.2 |  | 91.5861 | 10,195 |
|  | 1.1 |  | 114.9038 | 12,791 |
| 1,100 | 1.5 | 1,100 | 53.4698 | 5,952 |
|  | 1.4675 |  | 57.5088 | 6,402 |
|  | 1.4 |  | 67.2808 | 7,490 |
|  | 1.3 |  | 85.0740 | 9,470 |
|  | 1.2 |  | 105.7243 | 11,769 |
|  | 1.1 |  | 127.5217 | 14,195 |

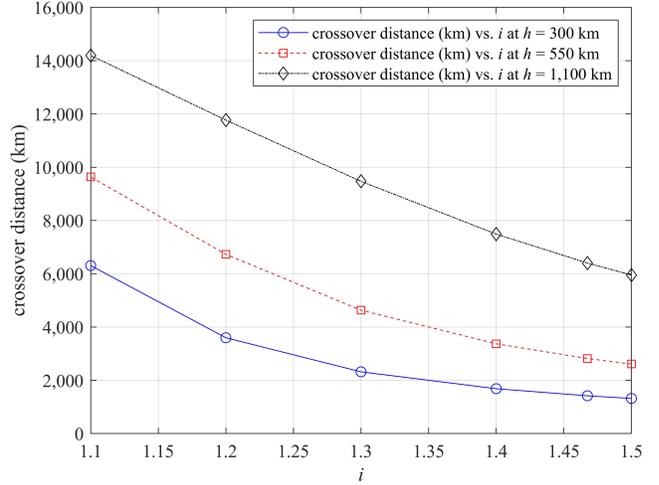

Fig. 3. Plot of crossover distance (km) for Scenario 1 vs. $i$ at different values of $h$ (km). As the value of optical fiber refractive index $i$ decreases for some value of satellite altitude $h$, the value of the crossover distance increases. This trend is similar for all values of $h$.

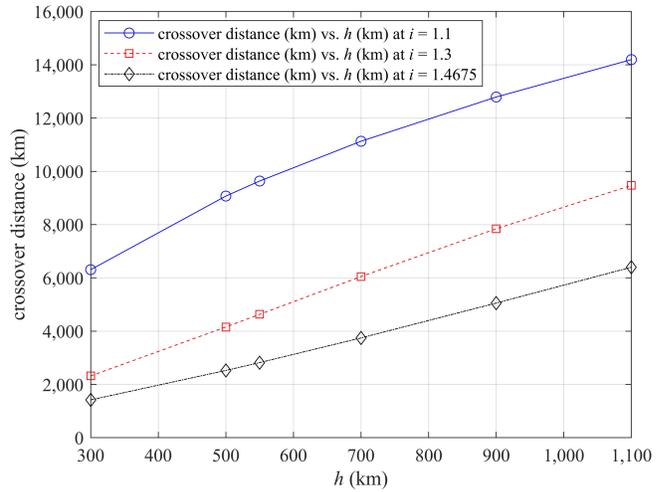

Fig. 4. Plot of crossover distance (km) for Scenario 1 vs. $h$ (km) at different values of $i$. As $h$ increases for some $i$, the crossover distance increases. A similar trend is seen for all values of $i$.

Also, $d_{XX'} = d_{Y'Y}$ and can be calculated using the Law of Cosines as

$$d_{XX'} = d_{Y'Y} = \sqrt{h^2 + r_{gs}^2 - 2hr_{gs}\cos\left(\alpha\left(\frac{\pi}{180}\right)\right)}, \quad (16)$$

where $\alpha = 90° - \varepsilon$. Here, note that $d_{XX'} = d_{Y'Y} \ll d_{X'Y'}$ and we assume the length of the chord $XY$ in this scenario to be equal to $d_{XY} = d_{XX'} + d_{X'Y'} + d_{Y'Y}$.

Substituting values of different distances in (14), we get (17). Finally, the crossover function for this scenario can be calculated as in (18). Substituting $r_{gs}$ in (18), we get (19).

### C. Scenario 3

In this scenario, which is depicted in Fig. 6, the satellite $X$ lies after $X'$ and satellite $Y$ lies before $Y'$, and the end-to-end propagation distance over the OWSN in this scenario is equal to

$$d_{\text{AB\_OWSN}} = d_{AX} + d_{X'Y'} - d_{X'X} - d_{YY'} + d_{YB}, \quad (20)$$

where

$$d_{X'X} = d_{YY'} = \sqrt{h^2 + r_{gs}^2 - 2hr_{gs}\cos\left(\alpha\left(\frac{\pi}{180}\right)\right)}. \quad (21)$$

Substituting values of different distances in (20), we get (22). The crossover function for this scenario is given by (23).

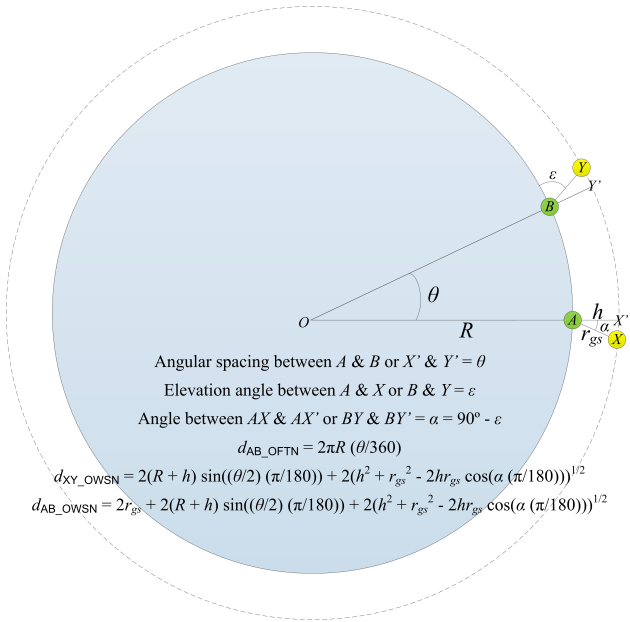

Fig. 5. Scenario 2 illustrating end-to-end propagation distance between $A$ and $B$ over the OWSN. The points $X'$ and $Y'$ represent the positions of the satellites in Scenario 1. The satellites $X$ and $Y$ are located before $X'$ and after $Y'$, respectively, in this scenario.

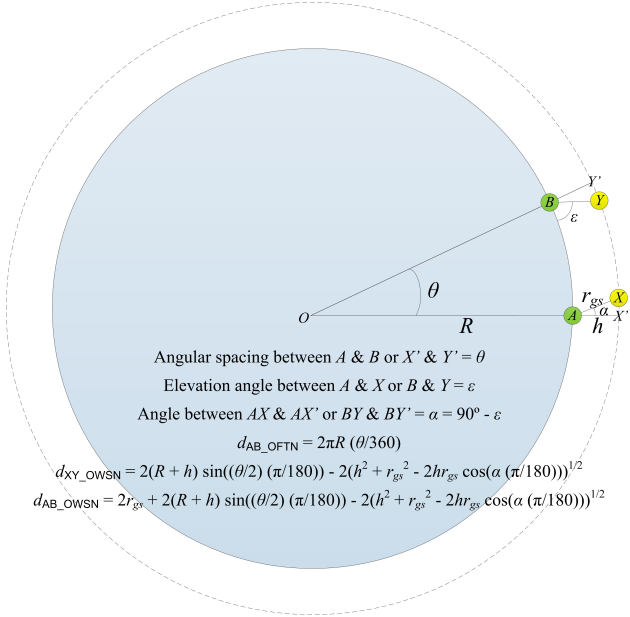

Fig. 6. Scenario 3 illustrating end-to-end propagation distance between $A$ and $B$ over the OWSN. The points $X'$ and $Y'$ represent the positions of the satellites in Scenario 1. The satellite $X$ lies after $X'$ and satellite $Y$ lies before $Y'$ in this scenario.

### D. Scenario 4

In this scenario, shown in Fig. 7, the LEO satellite $X$ is present after $X'$ while the LEO satellite $Y$ is present after $Y'$, and the end-to-end propagation distance over the OWSN in this scenario is

$$d_{AB\_OWSN} = d_{AX} + d_{X'Y'} - d_{X'X} + d_{Y'Y} + d_{YB}. \quad (24)$$

Since $d_{X'X} = d_{Y'Y}$, (24) simplifies to

$$d_{AB\_OWSN} = d_{AX} + d_{X'Y'} + d_{YB}. \quad (25)$$

Substituting values of different distances in (25), we get

$$d_{AB\_OWSN} = 2r_{gs} + 2(R+h)\sin\left(\left(\frac{\theta}{2}\right)\left(\frac{\pi}{180}\right)\right). \quad (26)$$

The crossover function for this scenario is given by (27). Note the similarity between the crossover functions of Scenario 1 and Scenario 4 in (13) and (27), respectively. However, in (13) in Scenario 1, $\varepsilon = 90°$ which means that $r_{gs}$ is equal to $h$ in this scenario whereas in (27) in Scenario 4, $\varepsilon$ is assumed equal to the minimum elevation angle for satellites in a constellation.

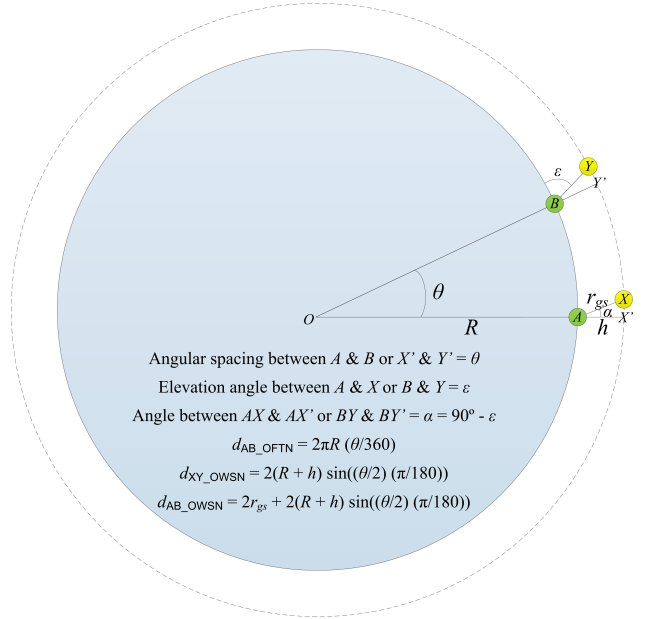

Fig. 7. Scenario 4 illustrating end-to-end propagation distance between $A$ and $B$ over the OWSN. The points $X'$ and $Y'$ represent the positions of the satellites in Scenario 1. The satellite $X$ is present after $X'$ while the satellite $Y$ is present after $Y'$ in this scenario.

The ground stations (also known as gateways) can only communicate with satellites above a certain minimum elevation angle $\varepsilon_{min}$, and it is specified as 25° for Phase I of Starlink in SpaceX's FCC filings [6]. We use $\varepsilon = 25°$ in Scenarios 2, 3, and 4 to calculate $r_{gs}$, crossover $\theta$, and crossover distance in these scenarios. The corresponding results are shown in Table 2. Note that for a certain value of $h$ and $i$ in Tables 1 and 2, the crossover distance is maximum for Scenario 2 and minimum for Scenario 3 since the end-to-end propagation distance between $A$ and $B$ over an OWSN is largest in Scenario 2 and smallest in Scenario 3. For example, for $h = 550$ km and $i = 1.1$ in these tables, the crossover distance for Scenario 1, Scenario 2, Scenario 3, and Scenario 4 is 9,636 km, 15,809 km, 4,981 km, and 12,507 km, respectively. This signifies that the crossover distance varies with the end-to-end propagation distance over an OWSN.

The corresponding results for the average crossover distance over all scenarios are given in Table 3. The last column in this

$$f_{1\_crossover}(\theta) = \frac{2\left(R\left[\sqrt{\left(\frac{R+h}{R}\right)^2 - \cos^2(\varepsilon)} - \sin(\varepsilon)\right]\right) + 2(R+h)\sin\left(\left(\frac{\theta}{2}\right)\left(\frac{\pi}{180}\right)\right)}{2\pi R\left(\frac{\theta}{360}\right)(i)} \quad (13)$$

$$d_{\text{AB\_OWSN}} = 2r_{gs} + 2(R+h)\sin\left(\left(\frac{\theta}{2}\right)\left(\frac{\pi}{180}\right)\right) + 2\left(\sqrt{h^2 + r_{gs}^2 - 2hr_{gs}\cos\left(\alpha\left(\frac{\pi}{180}\right)\right)}\right) \quad (17)$$

$$f_{2\_crossover}(\theta) = \frac{2r_{gs} + 2(R+h)\sin\left(\left(\frac{\theta}{2}\right)\left(\frac{\pi}{180}\right)\right) + 2\left(\sqrt{h^2 + r_{gs}^2 - 2hr_{gs}\cos\left(\alpha\left(\frac{\pi}{180}\right)\right)}\right)}{2\pi R\left(\frac{\theta}{360}\right)(i)} \quad (18)$$

$$f_{2\_crossover}(\theta) = \frac{\left\{\begin{array}{c} 2\left(R\left[\sqrt{\left(\frac{R+h}{R}\right)^2 - \cos^2(\varepsilon)} - \sin(\varepsilon)\right]\right) + 2(R+h)\sin\left(\left(\frac{\theta}{2}\right)\left(\frac{\pi}{180}\right)\right) + \\ 2\left(\sqrt{h^2 + \left(R\left[\sqrt{\left(\frac{R+h}{R}\right)^2 - \cos^2(\varepsilon)} - \sin(\varepsilon)\right]\right)^2 - 2h\left(R\left[\sqrt{\left(\frac{R+h}{R}\right)^2 - \cos^2(\varepsilon)} - \sin(\varepsilon)\right]\right)\cos\left(\alpha\left(\frac{\pi}{180}\right)\right)}\right) \end{array}\right\}}{2\pi R\left(\frac{\theta}{360}\right)(i)} \quad (19)$$

$$d_{\text{AB\_OWSN}} = 2r_{gs} + 2(R+h)\sin\left(\left(\frac{\theta}{2}\right)\left(\frac{\pi}{180}\right)\right) - 2\left(\sqrt{h^2 + r_{gs}^2 - 2hr_{gs}\cos\left(\alpha\left(\frac{\pi}{180}\right)\right)}\right) \quad (22)$$

$$f_{3\_crossover}(\theta) = \frac{\left\{\begin{array}{c} 2\left(R\left[\sqrt{\left(\frac{R+h}{R}\right)^2 - \cos^2(\varepsilon)} - \sin(\varepsilon)\right]\right) + 2(R+h)\sin\left(\left(\frac{\theta}{2}\right)\left(\frac{\pi}{180}\right)\right) - \\ 2\left(\sqrt{h^2 + \left(R\left[\sqrt{\left(\frac{R+h}{R}\right)^2 - \cos^2(\varepsilon)} - \sin(\varepsilon)\right]\right)^2 - 2h\left(R\left[\sqrt{\left(\frac{R+h}{R}\right)^2 - \cos^2(\varepsilon)} - \sin(\varepsilon)\right]\right)\cos\left(\alpha\left(\frac{\pi}{180}\right)\right)}\right) \end{array}\right\}}{2\pi R\left(\frac{\theta}{360}\right)(i)} \quad (23)$$

$$f_{4\_crossover}(\theta) = \frac{2\left(R\left[\sqrt{\left(\frac{R+h}{R}\right)^2 - \cos^2(\varepsilon)} - \sin(\varepsilon)\right]\right) + 2(R+h)\sin\left(\left(\frac{\theta}{2}\right)\left(\frac{\pi}{180}\right)\right)}{2\pi R\left(\frac{\theta}{360}\right)(i)} \quad (27)$$

TABLE 2
DIFFERENT SCENARIOS – CROSSOVER $\theta$ AND CROSSOVER DISTANCE FOR DIFFERENT VALUES OF $h$ AND $i$; $\varepsilon = 25°$ (I.E., $\alpha = 90° - \varepsilon$).

| $h$ (km) | $i$ | $r_{gs}$ (km) | Scenario 2 | | Scenario 3 | | Scenario 4 | |
|---|---|---|---|---|---|---|---|---|
| | | | $\theta_{crossover}$ (degrees) | $d_{crossover}$ (km) | $\theta_{crossover}$ (degrees) | $d_{crossover}$ (km) | $\theta_{crossover}$ (degrees) | $d_{crossover}$ (km) |
| 300 | 1.5 | 649 | 46.1906 | 5,141 | 2.3870 | 265 | 25.2535 | 2,811 |
| | 1.4675 | | 49.1359 | 5,469 | 2.5716 | 286 | 27.0867 | 3,015 |
| | 1.4 | | 56.3166 | 6,269 | 3.0628 | 340 | 31.8062 | 3,540 |
| | 1.3 | | 70.0930 | 7,802 | 4.2709 | 475 | 42.1522 | 4,692 |
| | 1.2 | | 87.9298 | 9,788 | 7.0397 | 783 | 58.7719 | 6,542 |
| | 1.1 | | 109.0009 | 12,134 | 18.7669 | 2,089 | 82.5393 | 9,188 |
| 500 | 1.5 | 1,032 | 72.0282 | 8,018 | 4.0254 | 402 | 41.6482 | 4,636 |
| | 1.4675 | | 75.8936 | 8,448 | 4.3609 | 485 | 44.5585 | 4,960 |
| | 1.4 | | 84.7855 | 9,438 | 5.2735 | 587 | 51.7962 | 5,765 |
| | 1.3 | | 100.0770 | 11,140 | 7.6348 | 849 | 66.1116 | 7,359 |
| | 1.2 | | 117.5468 | 13,085 | 13.6754 | 1,522 | 85.0675 | 9,469 |
| | 1.1 | | 136.5127 | 15,196 | 39.5710 | 4,404 | 107.3817 | 11,953 |
| 550 | 1.5 | 1,123 | 77.7858 | 8,658 | 4.4469 | 495 | 45.6406 | 5,080 |
| | 1.4675 | | 81.7732 | 9,102 | 4.8250 | 537 | 48.7809 | 5,430 |
| | 1.4 | | 90.8551 | 10,114 | 5.8596 | 652 | 56.4986 | 6,289 |
| | 1.3 | | 106.2106 | 11,823 | 8.5729 | 954 | 71.3952 | 7,947 |
| | 1.2 | | 123.4653 | 13,744 | 15.7126 | 1,749 | 90.4590 | 10,070 |
| | 1.1 | | 142.0222 | 15,809 | 44.7475 | 4,981 | 112.3582 | 12,507 |
| 700 | 1.5 | 1,389 | 93.5624 | 10,415 | 5.7532 | 575 | 57.2647 | 6,374 |
| | 1.4675 | | 97.7698 | 10,883 | 6.2736 | 698 | 60.9731 | 6,787 |
| | 1.4 | | 107.1297 | 11,925 | 7.7218 | 859 | 69.7980 | 7,769 |
| | 1.3 | | 122.3882 | 13,624 | 11.6972 | 1,302 | 85.7696 | 9,547 |
| | 1.2 | | 138.9859 | 15,472 | 23.0123 | 2,561 | 104.7135 | 11,656 |
| | 1.1 | | 156.5317 | 17,425 | 58.7699 | 6,542 | 125.4435 | 13,964 |
| 900 | 1.5 | 1,727 | 111.6662 | 12,430 | 7.6188 | 848 | 71.7972 | 7,992 |
| | 1.4675 | | 115.9666 | 12,909 | 8.3708 | 931 | 75.9976 | 8,459 |
| | 1.4 | | 125.3375 | 13,952 | 10.5211 | 1,171 | 85.6506 | 9,534 |
| | 1.3 | | 140.1980 | 15,606 | 16.8177 | 1,872 | 102.0762 | 11,363 |
| | 1.2 | | 156.0126 | 17,367 | 35.5476 | 3,957 | 120.4220 | 13,405 |
| | 1.1 | | 172.5383 | 19,206 | 74.1914 | 8,258 | 139.8560 | 15,568 |
| 1,100 | 1.5 | 2,049 | 127.1727 | 14,156 | 9.6810 | 1,077 | 85.1195 | 9,475 |
| | 1.4675 | | 131.4531 | 14,633 | 10.7311 | 1,194 | 89.5840 | 9,972 |
| | 1.4 | | 140.6841 | 15,661 | 13.8224 | 1,538 | 99.5820 | 11,085 |
| | 1.3 | | 155.0975 | 17,265 | 23.4653 | 2,612 | 115.9216 | 12,904 |
| | 1.2 | | 170.2490 | 18,952 | 49.7480 | 5,537 | 133.5734 | 14,869 |
| | 1.1 | | 186.0006 | 20,705 | 86.8717 | 9,670 | 151.9652 | 16,916 |

table is the average crossover distance, which is the average of the crossover distances of all four scenarios. For example, for $h = 550$ km and $i = 1.1$ in this table, the corresponding average crossover distance is 10,733 km and it is the average of 9,636 km, 15,809 km, 4,981 km, and 12,507 km, which are the crossover distances for Scenario 1, Scenario 2, Scenario 3, and Scenario 4, respectively, at this $h$ and $i$ in Tables 1 and 2. While simulating an OWSN (as described later in Sections IV and V), we calculate shortest paths over the OWSN at all time slots. We use the average crossover distance over all scenarios in Table 3 to evaluate and explain the simulation results for the latency comparison of OFTNs and OWSNs in Section V. We consider this a reasonable approach since the end-to-end latency of a simulated OWSN is the average of the end-to-end latencies of the shortest paths over the OWSN at all time slots. An illustration of the crossover distance in different scenarios as well as the average crossover distance vs. $i$ is provided in Fig. 8 when $h = 550$ km.

## IV. METHODOLOGY FOR CALCULATING LATENCY OF AN OFTN AND AN OWSN

In this section, we describe in detail the different steps of our methodology for calculating the latency in an OFTN and an OWSN. The values used in this study of the different parameters for the three different OFTNs and the three different OWSNs are listed in Table 4.

### A. Optical Fiber Terrestrial Network

For the three OFTNs, we use three different refractive indices, $i_1$ for the first OFTN (or OFTN1), $i_2$ for the second OFTN (or OFTN2), and $i_3$ for the third OFTN (or OFTN3). We consider different refractive indices for OFTN1, OFTN2, and OFTN3 to study the impact of optical fiber refractive index on latency in OFTNs as well as for comparative analysis of latency between OFTNs and OWSNs. For example, we use 1.4675 as the value for $i_3$, which is the refractive index of a single-mode optical fiber (manufactured by Corning®) that is suitable for long-distance communications at 1,310 nm operating wavelength [12]. Technological developments may lead to a reduction of optical fiber refractive index in future and we assume lower values for $i_2$ and $i_1$ accordingly. The speed of light in optical fiber with a refractive index $i$ can be calculated using $c/i$, where the value of the speed of light in vacuum or $c$ is 299,792,458 m/s [30]. For example, we consider the speed of light in OFTN3 to be $c/1.4675$ or 204,287,876 m/s.

We calculate the latency of an inter-continental long-distance connection between two cities over an OFTN by dividing the shortest distance between two cities along the surface of the Earth with the speed of light in optical fiber in that OFTN. To calculate the shortest distance between two cities, we use their coordinates (latitudes and longitudes) on the surface of the Earth. To this end, we use the coordinates for the locations of the financial stock markets (i.e., New York Stock Exchange, Dublin Stock Exchange, Sao Paulo Stock Exchange, London Stock Exchange, Toronto Stock Exchange, and Sydney Stock Exchange) within these cities, and the shortest distance for New York–Dublin, Sao Paulo–London, and Toronto–Sydney inter-continental connections along Earth's surface is calculated as 5,121 km, 9,514 km, and 15,585 km, respectively. For this calculation, the radius of the Earth is considered as 6,378 km. For example, the latency of the shortest distance path for the Toronto–Sydney inter-continental connection over OFTN3 having a refractive index of 1.4675 and a speed of light in optical fiber of 204,287,876 m/s is calculated as 76.29 ms.

TABLE 3
AVERAGE CROSSOVER DISTANCE.

| $h$ (km) | $i$ | Average $d_{crossover}$ (km) |
|---|---|---|
| 300 | 1.5 | 2,384 |
| | 1.4675 | 2,548 |
| | 1.4 | 2,959 |
| | 1.3 | 3,822 |
| | 1.2 | 5,178 |
| | 1.1 | 7,430 |
| 500 | 1.5 | 3,849 |
| | 1.4675 | 4,105 |
| | 1.4 | 4,702 |
| | 1.3 | 5,876 |
| | 1.2 | 7,555 |
| | 1.1 | 10,156 |
| 550 | 1.5 | 4,211 |
| | 1.4675 | 4,472 |
| | 1.4 | 5,107 |
| | 1.3 | 6,339 |
| | 1.2 | 8,073 |
| | 1.1 | 10,733 |
| 700 | 1.5 | 5,208 |
| | 1.4675 | 5,529 |
| | 1.4 | 6,257 |
| | 1.3 | 7,631 |
| | 1.2 | 9,509 |
| | 1.1 | 12,265 |
| 900 | 1.5 | 6,488 |
| | 1.4675 | 6,839 |
| | 1.4 | 7,662 |
| | 1.3 | 9,171 |
| | 1.2 | 11,231 |
| | 1.1 | 13,956 |
| 1,100 | 1.5 | 7,665 |
| | 1.4675 | 8,050 |
| | 1.4 | 8,944 |
| | 1.3 | 10,563 |
| | 1.2 | 12,782 |
| | 1.1 | 15,372 |

TABLE 4
VALUES AND DESCRIPTIONS OF PARAMETERS FOR OFTNS AND OWSNS.

| Parameter | Value | Description |
|---|---|---|
| $i_1$ | 1.1 | Refractive index of optical fiber in OFTN1 |
| $i_2$ | 1.3 | Refractive index of optical fiber in OFTN2 |
| $i_3$ | 1.4675 | Refractive index of optical fiber in OFTN3 |
| $h_1$ | 300 km | Altitude of satellites in OWSN1 |
| $h_2$ | 550 km | Altitude of satellites in OWSN2 |
| $h_3$ | 1,100 km | Altitude of satellites in OWSN3 |
| $r_1$ | 3,400 km | LISL range of satellites in OWSN1 |
| $r_2$ | 5,016 km | LISL range of satellites in OWSN2 |
| $r_3$ | 7,540 km | LISL range of satellites in OWSN3 |
| $\varepsilon_{min}$ | 25° | Minimum elevation angle |
| $r_{gs1}$ | 649 km | Range of ground stations in OWSN1 |
| $r_{gs2}$ | 1,123 km | Range of ground stations in OWSN2 |
| $r_{gs3}$ | 2,049 km | Range of ground stations in OWSN3 |

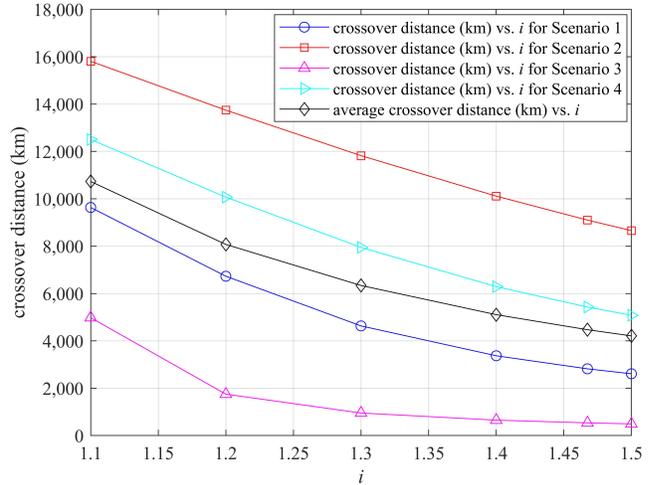

Fig. 8. Plot of crossover distance (km) for different scenarios as well as average crossover distance (km) vs. $i$ at $h$ = 550 km. For different values of $i$ in this figure, the crossover distance is maximum for Scenario 2 (depicted by the solid red line with square markers) and minimum for Scenario 3 (represented by the solid magenta line with triangle markers). The solid black line with diamond markers shows the average crossover distance. Similar trends exist at $h$ = 300 km and $h$ = 1,100 km and so the corresponding plots are omitted to avoid redundancy.

### B. Optical Wireless Satellite Network

To simulate an OWSN, we employ the satellite constellation for Phase I of Starlink and assume LISLs between satellites in this constellation. This constellation will consist of 1,584 LEO satellites in 24 orbital planes with 66 satellites per plane. The inclination of this constellation will be 53° and its altitude will be 550 km. We assume this constellation to be uniform, which means an equal spacing between orbital planes and an equal spacing between satellites within an orbital plane is assumed.

To study the impact of altitude of satellites on the latency of OWSNs as well as for the comparative analysis with OFTNs, we consider three different OWSNs based on this constellation at different altitudes. For the first OWSN (or OWSN1), $h_1$ is the altitude of its satellites and it is 300 km; for the second OWSN (or OWSN2), $h_2$ is the altitude of its satellites and it is 550 km; and for the third OWSN (or OWSN3), $h_3$ is the altitude of its satellites and it is 1,100 km. Note that $h_2$ is equal to the altitude of satellites in Starlink's Phase I constellation. We also consider OWSNs at a lower and a higher altitude as numerous upcoming satellite constellations are being planned at various VLEO and LEO altitudes.

The LISL range of a satellite is the distance over which it can establish LISLs with other satellites that are within its range. We consider the maximum LISL range for all satellites in an OWSN. The maximum LISL range of a satellite is a range that is only constrained by visibility. As described in [3], the maximum LISL range of satellites at 550 km altitude can be calculated as 5,016 km. Similarly, we calculate the maximum LISL range for satellites at 300 km and 1,100 km altitudes as 3,400 km and 7,540 km, respectively. We denote $r_1$, $r_2$, and $r_3$ as LISL ranges for satellites in OWSN1, OWSN2, and OWSN3, and set them to 3,400 km, 5,016 km, and 7,540 km, respectively. Using $\varepsilon = 25°$ (i.e., the minimum elevation angle specified for Phase I of Starlink in SpaceX's FCC filings [6]), $R = 6,378$ km, and $h = 300$ km, 550 km, and 1,100 km in (11), we set the range of ground stations in OWSN1, OWSN2, and OWSN3, i.e., $r_{gs1}$, $r_{gs2}$, and $r_{gs3}$ as 649 km, 1,123 km, and 2,049 km, respectively.

To calculate latency in an OWSN, we consider the speed of light in vacuum for LISLs and for laser links between ground stations and satellites. We identify all possible links in an OWSN, find their lengths, and calculate the latency of each link by dividing the length of a link by the speed of light in vacuum. Using Dijkstra's algorithm [31], we find the shortest path between two cities over an OWSN in terms of link latency; this is in fact the minimum-latency route between cities over that OWSN. The latency (or end-to-end latency) of the shortest path over an OWSN includes the latency of the laser link from the ground station in the source city on Earth to the ingress satellite of the OWSN in space, the latencies of the laser inter-satellite links in this path, and the latency of the laser link from the egress satellite of the OWSN in space to the ground station in the destination city on Earth.

We divide the time into time slots of one second duration, where a time slot represents a snapshot of an OWSN at that second. We find a shortest path (i.e., a route with minimum latency) at each time slot for an OWSN. We calculate the latency (or end-to-end latency) of an inter-continental connection over an OWSN by taking the average of the end-to-end latencies of all shortest paths at all time slots over the entire simulation duration. We run the simulation for a duration of one hour, i.e., 3,600 seconds or time slots. For example, the latency (or end-to-end latency) of OWSN1 for the New York–Dublin intercontinental connection is calculated as 18.84 ms, which is the average of the end-to-end latencies of the shortest paths at all 3,600 time slots.

## V. Latency Comparison – OFTNs vs. OWSNs

To study the impact on latency of optical fiber refractive index in OFTNs and altitude of satellites in OWSNs, we consider three different OFTNs and three different OWSNs as specified in Table 4. To compare these OFTNs and OWSNs in terms of latency, we examine them in three different inter-continental connection scenarios, including New York–Dublin, Sao Paulo–London, and Toronto–Sydney. We simulate the three different OWSNs using the well-known satellite constellation simulator STK Version 12.1 [32], the satellite constellation for Phase I of Starlink, and the parameters given in Table 4 and described in Section IV-B. For example, to simulate the OWSN1, we generate Starlink's Phase I constellation using $h_1$ as the altitude of satellites, $r_1$ as the LISL range of satellites, and $r_{gs1}$ as the range of ground stations; and we generate distinct IDs for the 1,584 satellites within the constellation.

After generating the satellite constellation corresponding to an OWSN and ground stations at locations of stock exchanges in various cities in STK, we extract the data of the OWSN from STK into Python, such as positions of satellites and ground stations, links between satellites, links between satellites and ground stations, and duration of the existence of these links. This data that is obtained from the STK simulator is discretized into time slots in Python. The discretized data includes all links that exist at a time slot as well as the positions of satellites at that time slot. Using this discretized data, we calculate the length and the latency for all links that exist at a time slot. Finally, we use the NetworkX library in Python [33] to find the shortest (or minimum-latency) path between ground stations in different cities over the OWSN at each time slot.

The orbital velocity $v_o$ of a satellite orbiting the Earth can be calculated using [34]

$$v_o = \sqrt{\frac{GM_E}{R+h}}, \quad (28)$$

where $G$ is the gravitational constant, $M_E$ is the mass of the Earth, $R$ is Earth's radius, and $h$ is the altitude of the satellite. Using $G = 6.673 \times 10^{-11}$ Nm$^2$/kg$^2$, $M_E = 5.98 \times 10^{24}$ kg, and $R = 6.378 \times 10^6$ m in (28), one can determine that the satellites in OWSN1 at 300 km altitude, OWSN2 at 550 km altitude, and OWSN3 at 1,100 km altitude travel at speeds of approximately 7.7 km/s, 7.6 km/s, and 7.3 km/s, respectively. Due to this high orbital speed of satellites in an OWSN, the ground station-to-ingress satellite link, satellite-to-satellite links, and egress satellite-to-ground station link or the latencies of these links change constantly. Consequently, the shortest path of an inter-continental connection over an OWSN between ground stations in two cities and/or its latency change at every time slot. As also stated earlier, we run the simulation of an OWSN for one hour or 3,600 time slots and find the shortest (or minimum-latency) path between ground stations in two cities over the OWSN at every time slot. For example, the shortest path calculated at the first time slot over OWSN3 for the Toronto–Sydney inter-continental connection is shown in Fig. 9. The shortest path is shown in yellow color in this figure while the satellites on this shortest path are shown in pink. This shortest path consists of the ground station at Toronto (located at the Toronto Stock Exchange), satellites *x12112* (ingress), *x10203* (intermediate hop), and *x10161* (egress) in the OWSN, and the ground station at Sydney (located at the Sydney Stock Exchange). In addition, the shortest distance path over an OFTN for the Toronto–Sydney inter-continental connection along Earth's surface is illustrated in this figure in green color.

Table 5 shows the latency (i.e., the end-to-end latency between source and destination points in two cities) of the three OFTNs and the three OWSNs for the three different inter-continental connection scenarios. Note that the latency of an inter-continental connection over an OWSN shown in this table is the average of the end-to-end latencies of the shortest paths that are found at all time slots. It is observed that the latency of OFTNs increases with the increase in the optical fiber refractive index in all scenarios. For example, the latency of OFTN1, OFTN2, and OFTN3 is 18.79 ms, 22.21 ms, and 25.07 ms, respectively, for the New York–Dublin connection. Recall that $i_1$ or the optical fiber refractive index for OFTN1, $i_2$ or the optical fiber refractive index for OFTN2, and $i_3$ or the optical fiber refractive index for OFTN3 are 1.1, 1.3, and 1.4675, respectively. The higher the optical fiber refractive index, the slower the speed of light through an OFTN, and the higher the latency.

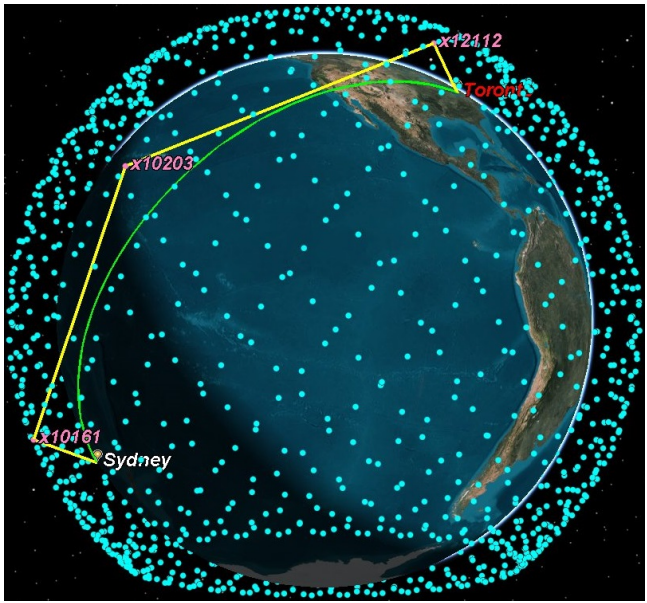

Fig. 9. The shortest path at first time slot for Toronto–Sydney inter-continental connection over OWSN3 is shown in yellow color and the satellites on this path are shown in pink. It consists of the ground station at Toronto Stock Exchange, satellites *x12112* (ingress), *x10203* (intermediate hop), and *x10161* (egress) in the OWSN, and the ground station at Sydney Stock Exchange. It's length is 18,201 km, and it's latency over OWSN3 having an altitude of 1,100 km is 60.71 ms. Furthermore, the shortest distance path for Toronto–Sydney inter-continental connection over an OFTN is shown in green color. This is the shortest path between Toronto and Sydney along the surface of the Earth, it's length is 15,585 km, and it's latency over OFTN3 having a refractive index of 1.4675 is 76.29 ms.

We also observe that the latency (or average latency) of OWSNs increases with the increase in the altitude of satellites in all scenarios. For example, the latency for the Sao Paulo–London connection is 34.13 ms, 35.31 ms, and 37.84 ms for OWSN1, OWSN2, and OWSN3, respectively. It should be noted that $h_1$, i.e., the altitude of satellites in OWSN1, $h_2$, i.e., the altitude of satellites in OWSN2, and $h_3$, i.e., the altitude of satellites in OWSN3 are 300 km, 550 km, and 1,100 km respectively. A higher altitude of satellites in an OWSN translates into longer uplink (i.e., the link between source ground station and ingress satellite), longer satellite-to-satellite links, and longer downlink (i.e., the link between egress satellite and destination ground station), which in turn translate into higher latency.

The results in Table 5 indicate that the latency of an OFTN or an OWSN is lowest for the New York–Dublin connection, increases for the Sao Paulo–London connection, and is highest for the Toronto–Sydney connection. For example, the latency of OFTN2 is 22.21 ms, 41.26 ms, and 67.58 ms, and the latency of OWSN2 is 19.58 ms, 35.31 ms, and 56.71 ms for the New York–Dublin, Sao Paulo–London, and Toronto–Sydney connections, respectively. One may recall that the shortest distance for New York–Dublin, Sao Paulo–London, and Toronto–Sydney connections along the Earth's surface is 5,121 km, 9,514 km, and 15,585 km, respectively. The longer the distance between two cities of an inter-continental connection, the higher the latency over an OFTN or an OWSN.

It is clearly seen from the results in Table 5 that all three OWSNs outperform OFTN2 and OFTN3 in terms of latency in all scenarios. For example, OWSN3 provides a latency improvement of 2.66%, 8.29%, and 10.30% for the New York–Dublin, Sao Paulo–London, and Toronto–Sydney connections, respectively, compared to OFTN2. The longer the inter-continental connection between a pair of cities, the greater the gain due to the higher speed of light over laser links in vacuum of space in OWSN3 compared to the speed of light over laser links in optical fiber in OFTN2, and the better the latency improvement offered by OWSN3 compared to OFTN2. The OWSN1 performs closely to OFTN1 for the New York–Dublin connection and outperforms it for Sao Paulo–London and Toronto–Sydney connections; OWSN2 performs better than OFTN1 for the Toronto–Sydney connection; and OWSN3 shows higher latency than OFTN1 in all scenarios.

As per Table 3, the average crossover distance is 7,430 km at $i = 1.1$ and $h = 300$ km, it is 10,733 km at $i = 1.1$ and $h = 550$ km, and it is 15,372 km at $i = 1.1$ and $h = 1,100$ km. These average crossover distances correspond to OFTN1 and OWSN1, OFTN1 and OWSN2, and OFTN1 and OWSN3, respectively. An OWSN can outperform an OFTN in terms of latency if the shortest terrestrial distance between two points on Earth is greater than the average crossover distance. For example, OWSN1 can outperform OFTN1 for communication between two points on Earth if the shortest distance between them is greater than 7,430 km. Note that the shortest distance between New York and Dublin along Earth's surface is 5,121 km. Since this distance is less than these three average crossover distances, OWSN1, OWSN2, and OWSN3 are not able to outperform OFTN1 for this connection. Similarly, the shortest terrestrial distance between Sao Paulo and London is 9,514 km, which is less than the average crossover distance of 10,733 km and 15,372 km. Hence, OWSN2 and OWSN3 cannot perform better than OFTN1 for the Sao Paulo–London connection.

Recall that the shortest terrestrial distance between Toronto and Sydney is 15,585 km, which is more than the average crossover distance of 15,372 km. However, OWSN3 exhibits a higher latency than OFTN1 for the Toronto–Sydney connection. Note that the average crossover distance of 15,372 km is calculated based on an OFTN having a refractive index of 1.1 and an OWSN at 1,100 km altitude where two satellites communicate directly without any intermediate hops or satellites. However, as discussed in Section IV-B, the LISL range of satellites (i.e., the maximum distance for any two satellites to communicate directly without any intermediate hops) is limited to 7,540 km in OWSN3. This LISL range leads to three hops or satellites on the first shortest path (i.e., the shortest path at first time slot) for the Toronto–Sydney connection over OWSN3 as shown in Fig. 9. Instead of a direct LISL between satellites *x12112* and *x10161*, the two satellites are connected through an intermediate hop or satellite *x10203* (resulting in two LISLs) due to the limitation on their LISL range, and this drives up the propagation distance to 18,201 km over this shortest path producing a latency of 60.71 ms for this shortest path over OWSN3 – a latency that is higher than that of OFTN1 for this connection. Similar to first shortest path over OWSN3 for the Toronto–Sydney connection, 3,500 shortest paths over this OWSN for this connection have three satellites or hops (or two LISLs) while 100 have four (or three LISLs), which translates into a higher average latency for OWSN3 compared to OFTN1 for this connection.

TABLE 5
LATENCY – OFTNs vs. OWSNs.

| Inter-Continental Connection | Latency (ms) | | | | | |
|---|---|---|---|---|---|---|
| | OFTN1 | OFTN2 | OFTN3 | OWSN1 | OWSN2 | OWSN3 |
| New York–Dublin | 18.79 | 22.21 | 25.07 | 18.84 | 19.58 | 21.62 |
| Sao Paulo–London | 34.91 | 41.26 | 46.57 | 34.13 | 35.31 | 37.84 |
| Toronto–Sydney | 57.18 | 67.58 | 76.29 | 55.34 | 56.71 | 60.62 |

## VI. CONCLUSIONS

A crossover function is proposed to enable the determination of the crossover distance for switching from an OFTN on Earth to an OWSN in space for lower latency data communications. The crossover distance depends upon the optical fiber refractive index $i$ in an OFTN and the altitude of satellites $h$ in an OWSN. The numerical results indicate that a higher $i$ and a lower $h$ result in a shorter crossover distance. The crossover function is examined in four different scenarios to account for the different end-to-end propagation distances that occur over the OWSN due to the orbital movement of satellites with time. It is observed from the numerical results that the crossover distance varies with the end-to-end propagation distance over an OWSN. It is minimum for Scenario 3 and maximum for Scenario 2 since the end-to-end propagation distance over an OWSN is smallest in Scenario 3 and largest in Scenario 2. The average crossover distance, which is the average of the crossover distances of all four scenarios, is calculated for different $h$ and $i$, and is used to evaluate the simulation results.

Furthermore, three different OFTNs having different $i$s and three different OWSNs with different $h$s are compared in terms of latency under three different scenarios for long-distance inter-continental data communications. The simulation results indicate that OWSN1 (i.e., the first OWSN with $h_1$ = 300 km), OWSN2 (i.e., the second OWSN with $h_2$ = 550 km) and OWSN3 (i.e., the third OWSN with $h_3$ = 1,100 km) outperform OFTN2 (i.e., the second OFTN with $i_2$ = 1.3) and OFTN3 (i.e., the third OFTN with $i_3$ = 1.4675) in all scenarios. For New York–Dublin connection, OWSN1, OWSN2, and OWSN3 perform better than OFTN2 by 15.17%, 11.84%, and 2.66%, respectively, while they provide an improvement in latency of 24.85%, 21.90%, and 13.76%, respectively, compared to OFTN3. For Sao Paulo–London connection, they perform better than OFTN2 by 17.28%, 14.42%, and 8.29%, respectively, and better than OFTN3 by 26.71%, 24.18%, and 18.75%, respectively. For Toronto–Sydney connection, they show an improvement of 18.11%, 16.08%, and 10.30%, respectively, compared to OFTN2 and an improvement of 27.46%, 25.67%, and 20.54%, respectively, compared to OFTN3.

Compared to OFTN1 (i.e., the first OFTN with $i_1$ = 1.1), OWSN1 performs better for the Sao Paulo–London and Toronto–Sydney connections by 2.33% and 3.22%, respectively, and OWSN2 performs better for the Toronto–Sydney connection by 0.82%. For the New York–Dublin connection, the corresponding average crossover distances are greater than the shortest terrestrial distance between New York and Dublin, and the three OWSNs are not able to offer lower latency than OFTN1 for this connection. Similarly, the corresponding average crossover distances are greater than the shortest terrestrial distance between Sao Paulo and London, and consequently OWSN2 and OWSN3 underperform for this connection compared to OFTN1. The OWSN3 shows a higher latency than OFTN1 for the Toronto–Sydney connection although the corresponding average crossover distance is smaller than the shortest terrestrial distance between Toronto and Sydney. The shortest or minimum-latency paths over OWSN3 for this connection mostly require three satellites (i.e., two LISLs) and sometimes four (i.e., three LISLs) resulting in a higher latency (or average latency) than OFTN1.

Furthermore, we observe for all scenarios that the latency of OWSNs increases as the altitude of satellites increases from 300 km to 1,100 km. Similarly, the latency of OFTNs increases with the increase in optical fiber refractive index from 1.1 to 1.4675. We also notice that the latency of an OFTN or an OWSN increases with the increase in length of the inter-continental connection from New York–Dublin to Toronto–Sydney. Compared to current OFTN comprising long-distance submarine optical fiber cables having a refractive index of 1.4675, OWSNs at all three altitudes show a significant improvement in latency. They can also offer better latency than a future OFTN with a refractive index of 1.3 in all scenarios. An OWSN at a lower altitude can even outperform a future OFTN with a refractive index of 1.1 for longer inter-continental connections. These findings identify OWSNs as a promising solution for HFT firms seeking revenue gains via improvement in latency of data communications among financial stock markets around the globe.

## VII. FUTURE CHALLENGES

In the following, we highlight some future challenges related to OWSNs and OFTNs that can arise from this work.

*A. Incorporating Processing Delay in the End-to-End Latency of OFTNs and OWSNs:*

For congestion-free OFTNs and OWSNs with very high data rate links, the queueing and transmission delays can be considered as negligible, and the end-to-end latency consists of processing and propagation delays. Therefore, in addition to the propagation delay, processing delay is an important part of the end-to-end latency that needs to be considered in both OWSN and OFTN. It is the delay that is incurred by a hop/node (i.e., an optical fiber relay station or a satellite) to process a packet, such as the time used to read the packet header to make appropriate routing and switching decisions, before sending the packet to the appropriate next hop. It depends upon the number of hops between the source and destination points (i.e., optical fiber relay stations or satellite ground stations) on the Earth's surface and becomes significant when the data communications has to go through several intermediate hops. The crossover function could be extended to incorporate this delay. It would also be interesting to study the effect of this delay on the end-to-end latency in OWSN and OFTN and to compare these two networks after incorporating this delay in the simulations.

*B. Making the Crossover Decision at Each Time Slot Based on Current Ingress and Egress Elevation Angles in the OWSN:*

Instead of comparing the average crossover distance and the shortest terrestrial distance, another approach to switch from the OFTN to the OWSN for long-distance lower latency data communications can be based on checking the elevation angles of the ingress and egress satellites at every time slot, calculating the corresponding crossover distance at a time slot, and comparing it with the shortest terrestrial distance between cities to make the crossover decision at that time slot. For instance, the crossover function for Scenario 2 in this case can be written as in (29). It would be interesting to evaluate such an approach in future.

*C. Incorporating Extra Distance to Account for the Zig-Zag Path of OFTNs:*

In this work, we consider the shortest distance between two cities over the OFTN along the Earth's surface. In reality, long-haul submarine optical fiber cables do not adhere to the shortest path to connect two points on Earth's surface and are installed along paths that avoid earthquake prone areas and difficult seabed terrains with high slopes. Another approach to model the shortest distance between cities over the OFTN can be to add an extra distance to account for the extra length of the long-haul submarine optical fiber cables due to the zig-zag nature of their path. This extra distance can be added to the shortest distance as a percentage of the shortest distance. For example, the crossover function for Scenario 2 in (29) can be calculated in this case as in (30). Such an approach could be investigated in future.

*D. Incorporating LISL Setup Delay in the End-to-End Latency of OWSNs:*

Due to pointing and acquisition during the formation of a LISL between a pair of satellites equipped with laser communication terminals, the delay to set up a LISL (which can also be referred to as the LISL setup delay) can be another significant component of the end-to-end latency in OWSNs. This delay is incurred whenever new LISLs are required to be established between pairs of satellites when there is a change in the shortest path between source and destination ground stations and one or more new satellites are introduced in the path requiring creation of new LISLs. Currently, this delay can vary from a few seconds to tens of seconds, e.g., Mynaric's CONDOR laser communication terminal needs approximately 30 seconds to establish LISLs between pairs of satellites in the OWSN for the first time but once the position and altitude of the satellites are exchanged, this delay reduces to 2 seconds [35]. This delay could be incorporated in the end-to-end latency of the shortest paths over the OWSN for a more accurate comparison of OWSN and OFTN in terms of latency. The current LISL setup times are prohibitive and in next-generation OWSNs (that may become fully operational by mid to late 2020s), a satellite will be limited to set up only permanent LISLs [3] with neighboring satellites that are always within its LISL range. Our work aims at next-next-generation OWSNs that may come into existence in early to mid-2030s where we envisage LISL setup times in milliseconds due to advancements in satellite laser communication terminal's pointing, acquisition, and tracking technology, which will enable a satellite to instantaneously set up an LISL with any neighboring satellite that is currently within its LISL range.

*E. Investigating the Effect of Different LISL Ranges on the End-to-End Latency of OWSNs:*

In this work, we consider the maximum LISL range for all satellites in an OWSN, where the maximum LISL range of a satellite is a range that is only constrained by visibility. Different LISL ranges are likely to impact latency of an OWSN differently. It has been concluded that the number of neighbors of a satellite increases with the increase in LISL range of satellites [3]. The LISL range may affect the connectivity of satellites within the OWSN. This may impact the shortest paths over the OWSN for long-distance inter-continental data communications between ground stations in different cities, and this may influence the latency of the OWSN. It would be interesting to investigate the effect of different LISL ranges on the latency (or end-to-end latency) of an OWSN.

ACKNOWLEDGMENT

This work has been supported by the National Research Council Canada's (NRC) High Throughput Secure Networks program (CSTIP Grant #CH-HTSN-625) within the Optical Satellite Communications Consortium Canada (OSC) framework. The authors would like to thank AGI for the STK platform.

$$f_{2\_crossover}(\theta) =$$

$$\frac{\left\{\begin{array}{l}\left(R\left[\sqrt{\left(\frac{R+h}{R}\right)^2 - \cos^2(\varepsilon_1)} - \sin(\varepsilon_1)\right]\right) + \left(R\left[\sqrt{\left(\frac{R+h}{R}\right)^2 - \cos^2(\varepsilon_2)} - \sin(\varepsilon_2)\right]\right) + 2(R+h)\sin\left(\left(\frac{\theta}{2}\right)\left(\frac{\pi}{180}\right)\right) + \\ \left(\sqrt{h^2 + \left(R\left[\sqrt{\left(\frac{R+h}{R}\right)^2 - \cos^2(\varepsilon_1)} - \sin(\varepsilon_1)\right]\right)^2 - 2h\left(R\left[\sqrt{\left(\frac{R+h}{R}\right)^2 - \cos^2(\varepsilon_1)} - \sin(\varepsilon_1)\right]\right)\cos\left(\alpha_1\left(\frac{\pi}{180}\right)\right)}\right) + \\ \left(\sqrt{h^2 + \left(R\left[\sqrt{\left(\frac{R+h}{R}\right)^2 - \cos^2(\varepsilon_2)} - \sin(\varepsilon_2)\right]\right)^2 - 2h\left(R\left[\sqrt{\left(\frac{R+h}{R}\right)^2 - \cos^2(\varepsilon_2)} - \sin(\varepsilon_2)\right]\right)\cos\left(\alpha_2\left(\frac{\pi}{180}\right)\right)}\right)\end{array}\right\}}{2\pi R\left(\frac{\theta}{360}\right)(i)}, \quad (29)$$

where $\varepsilon_1$ and $\varepsilon_2$ are the elevation angles for satellites *X* and *Y*, respectively, $\alpha_1 = 90° - \varepsilon_1$, and $\alpha_2 = 90° - \varepsilon_2$.

$$f_{2\_crossover}(\theta) =$$

$$\frac{\left\{\begin{array}{l}\left(R\left[\sqrt{\left(\frac{R+h}{R}\right)^2 - \cos^2(\varepsilon_1)} - \sin(\varepsilon_1)\right]\right) + \left(R\left[\sqrt{\left(\frac{R+h}{R}\right)^2 - \cos^2(\varepsilon_2)} - \sin(\varepsilon_2)\right]\right) + 2(R+h)\sin\left(\left(\frac{\theta}{2}\right)\left(\frac{\pi}{180}\right)\right) + \\ \left(\sqrt{h^2 + \left(R\left[\sqrt{\left(\frac{R+h}{R}\right)^2 - \cos^2(\varepsilon_1)} - \sin(\varepsilon_1)\right]\right)^2 - 2h\left(R\left[\sqrt{\left(\frac{R+h}{R}\right)^2 - \cos^2(\varepsilon_1)} - \sin(\varepsilon_1)\right]\right)\cos\left(\alpha_1\left(\frac{\pi}{180}\right)\right)}\right) + \\ \left(\sqrt{h^2 + \left(R\left[\sqrt{\left(\frac{R+h}{R}\right)^2 - \cos^2(\varepsilon_2)} - \sin(\varepsilon_2)\right]\right)^2 - 2h\left(R\left[\sqrt{\left(\frac{R+h}{R}\right)^2 - \cos^2(\varepsilon_2)} - \sin(\varepsilon_2)\right]\right)\cos\left(\alpha_2\left(\frac{\pi}{180}\right)\right)}\right)\end{array}\right\}}{\left(2\pi R\left(\frac{\theta}{360}\right) + \Delta\left(2\pi R\left(\frac{\theta}{360}\right)\right)\right)(i)}, \quad (30)$$

where $\Delta$ can be assumed as 0.1, for instance, to account for 10% extra distance due to the zig-zag path of long-haul submarine optical fiber cables.

TABLE 6
LIST OF ABBREVIATIONS.

| Abbreviation | Description |
| --- | --- |
| HFT | High-frequency trading |
| HydRON | High thRoughput Optical Network |
| LEO | Low Earth orbit |
| LISL | Laser inter-satellite link |
| OFTN | Optical fiber terrestrial network |
| OFTN1 | First optical fiber terrestrial network |
| OFTN2 | Second optical fiber terrestrial network |
| OFTN3 | Third optical fiber terrestrial network |
| OWSN | Optical wireless satellite network |
| OWSN1 | First optical wireless satellite network |
| OWSN2 | Second optical wireless satellite network |
| OWSN3 | Third optical wireless satellite network |
| VLEO | Very low Earth orbit |

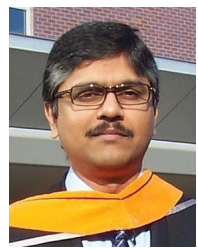

**Aizaz U. Chaudhry** (M'10–SM'20) received the B.Sc. degree in Electrical Engineering from the University of Engineering and Technology Lahore in 1999. He received the M.A.Sc. and Ph.D. degrees in Electrical and Computer Engineering from Carleton University in 2010 and 2015, respectively. He is a Senior Research Associate with the Department of Systems and Computer Engineering at Carleton University. Previously, he worked as an NSERC Postdoctoral Research Fellow at Communications Research Centre Canada. His research interests include the application of machine learning and optimization in wireless networks. His research work has been published in refereed venues, and has received several citations. He has authored and co-authored more than thirty publications. He is a licensed Professional Engineer in the Province of Ontario, a Senior Member of IEEE, and a Member of IEEE ComSoc's Technical Committee on Satellite and Space Communications.

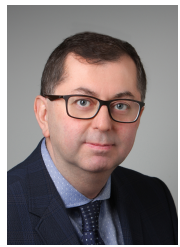

**Halim Yanikomeroglu** (F'17) is a Full Professor at Carleton University. He received his Ph.D. from University of Toronto in 1998. He contributed to 4G/5G technologies and non-terrestrial networks. His industrial collaborations resulted in 39 granted patents. He supervised or hosted in his lab a total of 140 postgraduate researchers. He co-authored IEEE papers with faculty members in 80+ universities in 25 countries. He is a Fellow of IEEE, Engineering Institute of Canada, and Canadian Academy of Engineering, and an IEEE Distinguished Speaker for ComSoc and VTS. He is currently chairing the WCNC Steering Committee, and he is a Member of PIMRC Steering Committee and ComSoc Emerging Technologies Committee. He served as the General Chair of two VTCs and TP Chair of three WCNCs. He chaired ComSoc's Technical Committee on Personal Communications. He received several awards including ComSoc Wireless Communications TC Recognition Award (2018), VTS Stuart Meyer Memorial Award (2020), and ComSoc Fred W. Ellersick Prize (2021).